\pdfoutput=1
\RequirePackage{fix-cm}
\documentclass[twocolumn]{svjour3}          % twocolumn
\smartqed  % flush right qed marks, e.g. at end of proof
\usepackage{graphicx}
\usepackage{scrtime}
\usepackage[dont-mess-around]{fnpct}

\usepackage{etoolbox}
\newcommand*{\affaddr}[1]{#1} % No op here. Customize it for different styles.
\newcommand*{\affmark}[1][*]{\textsuperscript{#1}}
\usepackage{amsmath}
\usepackage{listings}
\usepackage[toc,page]{appendix}
\usepackage{enumitem}
\usepackage{hyperref}
\setlist{nosep}

\journalname{Computing and Software for Big Science}
\begin{document}

\title{Dynamo -- Handling Scientific Data Across Sites and Storage Media}
%\titlenote{Managing data dynamically with a rich policy language.}
\author{Yutaro Iiyama\protect\affmark[1] \and Benedikt Maier\affmark[2] \and Daniel Abercrombie\affmark[2] \and Maxim Goncharov\affmark[2] \and Christoph Paus\affmark[2]}
\authorrunning{Y.~Iiyama \and B.~Maier \and D.~Abercrombie \and M.~Goncharov \and C.~Paus}
\institute{
  \affaddr{\affmark[1] International Center for Elementary Particle Physics, The University of Tokyo, Tokyo, Japan} \\
  \affaddr{\affmark[2] Laboratory for Nuclear Science, Massachusetts Institute of Technology, Cambridge, MA 02139, USA} \\
}

\date{Received: date / Accepted: date}
% The correct dates will be entered by the editor

\maketitle

\begin{abstract}
Dynamo is a full-stack software solution for scientific data management. Dynamo's architecture is modular, extensible, and customizable, making the software suitable for managing data in a wide range of installation scales, from a few terabytes stored at a single location to hundreds of petabytes distributed across a worldwide computing grid. This article documents the core system design of Dynamo and describes the applications that implement various data management tasks. A brief report is also given on the operational experiences of the system at the CMS experiment at the CERN Large Hadron Collider and at a small scale analysis facility.
% \PACS{PACS code1 \and PACS code2 \and more}
% \subclass{MSC code1 \and MSC code2 \and more}
\end{abstract}

\section{Introduction}

Facilities like the Large Hadron Collider (LHC)~\cite{lhc} at CERN, Geneva, with its experiments including Alice~\cite{alice}, ATLAS~\cite{atlas}, CMS~\cite{cms}, and LHCb~\cite{lhcb} are giving rise to very large amounts of experimental data that is now already close to an exabyte and will continue to grow substantially in the next two decades. Thousands of scientists around the globe are analyzing these data in the pursuit of finding evidence for new physics phenomena that are not predicted by the established theories. Often, scientific results are produced just in time for the next conference. In such a fast-paced environment at the cutting edge of research, one of the key challenges the collaborations are confronted with is the efficient and reliable management of their data that are being taken and analyzed by a large number of collaborators. This is especially important given the fact that the experimental data are the core asset at the center of multi-billion dollar projects like the LHC.

The moment we accumulate data of a large volume, the question of how to do data management arises. Even with this problem being a very old and well-studied one~\cite{phedex,diraclhcb,jalien,rucio}, no universal solution or implementation has emerged. The reason is that data management has to address the specific set of requirements of the given environment, such as the preexisting data organization concepts and the structure of distributed computing infrastructure. Those factors have a strong influence on the design of data management products.

In the case of the LHC experiments, one of the defining constraints is the distribution and types of the available data storage. Specifically, in the tiered computing approach taken by the LHC experiments~\cite{alice,cms,lhcb}, in which in the order of 100 geographically separated sites provide data storage, sites are heterogeneous in terms of capacity, mass storage technology, network interconnectivity, level of support, etc. For example, in CMS, Tier-1 sites provide large archival tape storage, while disk storage is provided by both the Tier-1 and smaller and more copious Tier-2 sites. Data on tape systems are not immediately accessible (cold storage), but disk pools, which allow immediate read and write access, are limited in capacity. Thus, nontrivial decisions have to be taken on which pieces of data to keep on disk, with how many copies, and where.

Another important factor in designing a data management product is how the data are actually utilized. Data usage in the experiments can be categorized into two big classes: production access, made by data processing tasks planned by the experimental collaboration to produce collaboration-wide common datasets\footnote{Data in CMS are mainly organized in datasets, which are collections of files sharing semantic properties (See Section~\ref{subsec:concepts}).}, and user analysis access, made by individual analysts. The biggest difference between the two classes is that production access is predictable while user analysis is inherently unpredictable. As an example, the reprocessing of the data with updated calibrations is carefully planned and the necessary inputs can be staged from tape without injecting any additional latency into the reprocessing schedule. On the other hand, a user might one day decide to analyze a dataset that has not been accessed for multiple years, or hundreds of users might want to read the same dataset at the same time. To avoid bottlenecks in the analysis tasks, data management must provide some slack in the form of distributed copies of datasets, and possess certain intelligence to keep that slack under control.

The initial approach in CMS towards data management was to ensure that datasets can be efficiently and safely transferred from one storage site to another, with a rich set of permissions to identify who is allowed to perform certain actions on the data. Sites were put in charge to install local software agents to execute transfers and communicate with the central agents about their progress. The intelligence about which data was supposed to be available at the sites was to be provided by data managers, who were individuals appointed by the subgroups of the collaboration. Each subgroup was assigned three to five specific Tier-2 sites to fill with the datasets of their interests, with exclusive ownership (custodianship) on these datasets. Some coordination was required to decide who was in charge of the large datasets, such as ones containing at least one muon as determined by the CMS trigger system, because they are used by almost all physics analysis groups. This coordination of data ownership was quite time-consuming in certain cases.

For the first few years this concept worked, because there was enough disk space, a lot of interest and support from the sites and the data managers, and there were relatively few datasets. Over time, sites and data managers had less resources, and with the rapidly growing amount of data and number of datasets, the scheme became virtually unmanageable. Another important development was that the strict rule of re-processing the detector and Monte Carlo simulation data only at the Tier-1 sites proved to be a major bottleneck on the production process. Moving the re-processing also to the Tier-2 sites meant a substantial increase in data transfers, which became impossible to support with the team available for computing operations. In short, there was a large need for automation and intelligence, which was particularly evident in the computing operations community in CMS at the time.
% do we really want to mention the reprocesing issue?

Studying this situation, there were a number of key conclusions reached.
\begin{itemize}
\item Users should not have to care where their analysis jobs run, as long as they finish successfully and quickly;
\item subgroups did not want to and could not manage their own data;
\item sites did not want to manage the exact data content of their storage; and
\item data production systems needed an automatic way to spread the data across all production sites with the least amount of effort.
\end{itemize}

To address these points, we introduced an automated data management system, which we dubbed Dynamo. Dynamo was developed with the goal of eliminating or at least minimizing human interactions with the data management system, and at the same time optimizing the way the storage is used to hold the data for user analysis and for the data production system. In addition, a number of important simplifications and features were introduced to the data management model. To name a few:
\begin{itemize}
\item Sites were opened to any datasets that users or production were interested in.
\item Data ownership by subgroups was deprecated, and was replaced with that by two main groups: Analysis and Production.
\item Predefined data replication and deletion rules (\textit{data placement policies}) were introduced to fill the disk space automatically with popular data replicas, while removing less popular data replicas.
\item A fully automatized site consistency enforcement was introduced to address any failures in the data management system.
\item A fully automatic site evacuation was introduced to quickly and efficiently deal with major site failures.
\item An interface to the batch submission system was provided to automatically download data that are only available on tape to disk, when required by the users.
\end{itemize}

Dynamo is a software package which enables intelligent and flexible data management by incorporating a rich representation of the global storage system and its contents. The package can be used as a high-level intelligence layer of a data management software stack, only making the data placement decisions but not performing the actual file transfers and deletions, or as a full-stack standalone data management product. The data placement policies in Dynamo are expressed in a human-readable syntax and can be easily defined at run time, increasing the transparency of data management to the collaborators while minimizing the necessary human intervention to the system.

In this article, we document the design and some implementation details of Dynamo. We then describe various Dynamo applications, which are the key components of the system that implement actual data management operations. Finally, we introduce real-world use cases of Dynamo in the CMS collaboration and at a single local analysis facility.

\section{Overview of the system}
\label{sec:overview}

\subsection{Basic functionalities and assumptions}
\label{subsec:basic_functionalities}

The core purpose of Dynamo is to oversee the physical placement of data and to orchestrate their bulk transfers among the storage sites. It is intended to operate in an ecosystem including a production system, which creates data files of experimental data and/or simulation, and an analysis system, which schedules and distributes analysis jobs. These systems interact with Dynamo through its REST~\cite{rest} application programming interface (API). However, the production and analysis systems are not absolute necessities for Dynamo's operation, and it is possible in small-scale installations that data management is performed manually by administrators issuing commands to Dynamo's REST API or by an appropriate set of Dynamo applications, which are explained later. It should be noted that end users, who perform the analyses over the data files, would normally not interact with Dynamo directly to e.g. download single files to their laptops, as there are other tools, such as XRootD~\cite{xrd}, to fulfill those needs.

It is also expected that there independently exists an authoritative catalog of all files and their groupings (datasets), presumably populated by the production system.  While Dynamo can in principle serve as such a catalog, it stores a minimal and fixed set of metadata of files and datasets by design, and would likely fall short of an experiment's needs. The authoritative catalog needs to provide some API to allow Dynamo to read its content, but does not need to be able to initiate interactions with Dynamo.

In a full-scale installation under the data-local paradigm (compute moves close to the data), Dynamo enters typical workflows in the following ways:
\begin{itemize}
    \item The production system schedules to process a source dataset and create a derived-format dataset. It first queries Dynamo to locate the source dataset at a disk storage site. It also requests Dynamo to create an additional copy of the dataset at another site to split the production work in half. After the production jobs are submitted and run at the two sites, a half of the derived dataset that is created locally is left at each site. The production system then requests Dynamo to copy the half on one of the sites to the other site, effectively consolidating the dataset and creating a full copy. Dynamo deletes the additional copy of the source dataset and the remaining half of the derived dataset after the workflow is completed.
    \item The analysis system accepts a task from a user. It first queries Dynamo to locate the dataset the user wants to analyze. If the dataset is found on a disk storage site, the analysis jobs are sent to the site. If it is only found on a tape archive, the analysis system requests Dynamo to create a copy of the dataset at a disk site (stage the dataset), and regularly queries it to check for completion of the transfer.
    \item Alternatively, although the dataset is found at a disk storage site, the analysis jobs need to run at a specific site because e.g. the site has a special hardware configuration. The analysis system requests a copy of the dataset at the site to Dynamo and waits for the completion of the transfer.
\end{itemize}

Clearly, what is a typical workflow depends largely on the scale of the installation and various constraints of the experiment, such as the available disk storage and the connectivity of the sites. Dynamo is designed to be flexible enough to accommodate vastly different use cases. In the next sections, we describe its design and the underlying components.

\subsection{System design}
\label{subsec:system_design}

Dynamo is written in Python~\cite{python} 2.7 with a modular architecture. The central component depends only on the Python standard library, to decouple the system core from specific technologies for storage, transfer, and metadata management. Interface to various external services and to internal bookkeeping persistence are provided as plugins. A minimum set of plugins required to perform the standard tasks in a small-scale standalone environment are packaged together with the core software.

\begin{figure*}
  \includegraphics[width=0.975\textwidth]{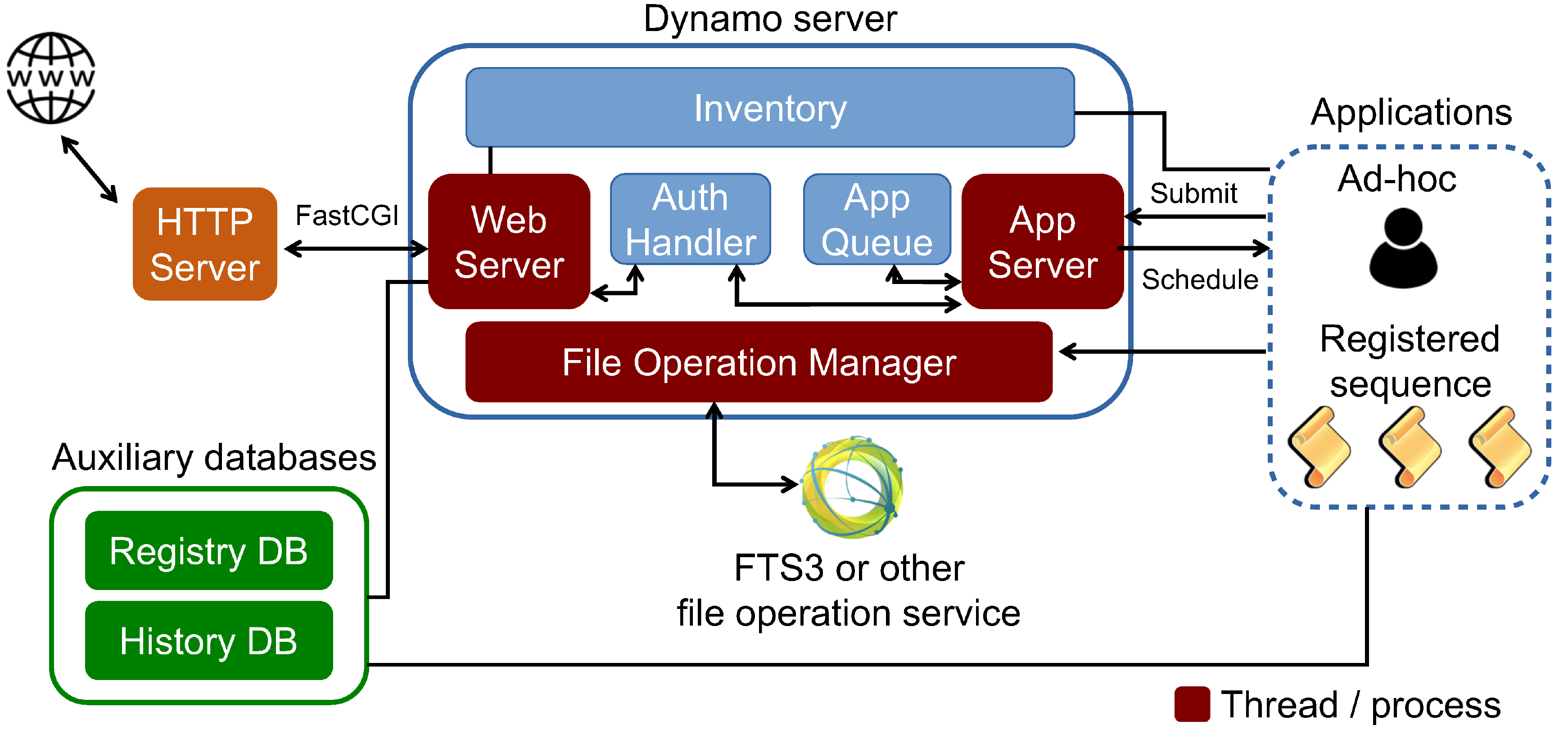}
  \caption{Schematic of the Dynamo system.}
  \label{fig:system_schematic}
\end{figure*}

A schematic of the Dynamo system is shown in Figure~\ref{fig:system_schematic}. The core of the system is the Dynamo server process, which possesses the full image of the global storage system under its management, called the \textit{inventory}. In the inventory, sites, datasets, and other entities described in Section~\ref{subsec:concepts} are all represented as objects interlinked with each other through one-to-one or one-to-many references (See Fig.~\ref{fig:inventory}). The entirety of the inventory, excluding the information on individual files, are kept in RAM, allowing fast execution of flexible and complex data placement algorithms. Information on files are not required until the data placement decisions are taken and when actual transfers and deletions take place, at which point it is automatically and transiently loaded from a persistence provider into the inventory image. Note that everything beyond the data placement decisions can be outsourced to external systems through plugins; by doing so, it is possible to operate Dynamo without any file-level information. Each object in RAM has an average footprint of a few hundred bytes. As explained later, the inventory for the CMS experiment, which is arguably one of the largest use cases, requires approximately 8 gigabytes of RAM.

The server process is meant to stay running continuously, but if it is to be stopped and restarted, the inventory image in RAM needs to be somehow persisted. In principle, persistence can be provided in any form with an appropriate plugin; one can even choose to serialize the image into an ASCII text file if desired. In practice, a relational database is used as the persistence provider, with a table for each class of objects in the memory. This is because it is more desirable to take real-time backup of the inventory at each update than just at the system stop, and relational databases naturally support frequent data insertion and update operations. However, it should be stressed that Dynamo is not a database-centric application, which is its main distinction to be drawn with respect to existing data management solutions.

Individual data management tasks, such as identifying data units to be copied or deleted and initiating the file operations, are carried out by Dynamo \textit{applications}, which are child processes of the server process. As child processes, applications inherit the inventory image from the server. Owing to the design that decouples the persistence from the inventory content, applications are written as simple Python programs that have access to the inventory as a normal Python object. Thus, at its core, from a technical perspective, Dynamo is an engine to execute an arbitrary Python program with one specific large structure (inventory) pre-loaded in memory. Default applications for standard tasks are included in the software package, and are described in Section~\ref{sec:applications}.

Because an application is a child process of the server, any modifications it makes to the inventory within its address space are discarded automatically at the end of its execution, and are not visible from the server or the other applications that may be running concurrently. However, pre-authorized applications can communicate the changes they make to the inventory back to the server before the process termination. Such applications are said to be write-enabled.

Although not a component of the Dynamo server, MySQL~\cite{mysql} database schemas for two auxiliary databases, \textit{registry} and \textit{history}, are included in the software package. These databases are used by the default applications and web modules for asynchronous inter-process communications (the registry database) and recording operation histories (the history database).

If an application performs actual file transfer and / or deletion operations, it can do so using any means (arbitrary Python programs can become Dynamo applications), but can also choose to delegate the operations to the built-in \textit{file operation manager} (\texttt{fom}). File operation manager is one of the default Dynamo applications that communicates with FTS3~\cite{Ayllon_2014} or other file operation services and updates the inventory as file transfers and deletions complete. File transfer and deletion commands are passed to the file operation manager through the registry database. The \texttt{fom} application is described in detail in Section~\ref{subsec:fom}.

Within the Dynamo server process, the \textit{app server} component runs in a dedicated thread and listens for requests to execute applications. Execution requests come from a command-line client or the built-in \textit{application scheduler}. The command-line client is mainly used by application developers and system administrators who need ad-hoc access to the inventory. The application scheduler should be used in the production environment, where fixed sets of applications are executed repeatedly. The scheduler makes execution requests of sequences of registered applications with configurable intervals between the requests. The app server component authenticates and authorizes the user (when the command-line client is used) and the application (if write access to the inventory is requested). The applications are then placed in a queue, from which they are picked up in the order of arrival by the server.

The Dynamo server can also run a web server as a child process. The web server communicates with an external HTTP(S) server through FastCGI~\cite{fastcgi} and exports a web page (HTML document) or provides a REST interface, depending on the requested URL. External systems (e.g. the production system) normally interact with Dynamo through the web interface. The actual content delivered through the web server is created by web server modules, which are easily expandable according to the needs. Default web modules for basic tasks are included in the Dynamo software package. Many modules are accessible only by authorized users. The web server uses the same authentication and authorization mechanism as the app server component of the Dynamo server. In a manner similar to the applications, web modules can be either read-only or write-enabled.

Finally, Dynamo server processes can operate concurrently on multiple machines communicating with each other. Each server instance in such a setup is equipped with an inventory image and optionally a local persistence provider. With the use of a load-balancing mechanism such as \texttt{keepalived}~\cite{keepalived}, linked parallel Dynamo instances can share the tasks of running applications and responding to HTTP requests. The multi-server setup also provides resiliency against individual node failures. As detailed in Section~\ref{subsec:parallelization}, the protocol for communications between the servers is equipped with a locking mechanism to ensure consistency of the copies of the inventory in the system.

\subsection{Concepts}
\label{subsec:concepts}

\begin{figure*}
  \includegraphics[width=0.975\textwidth]{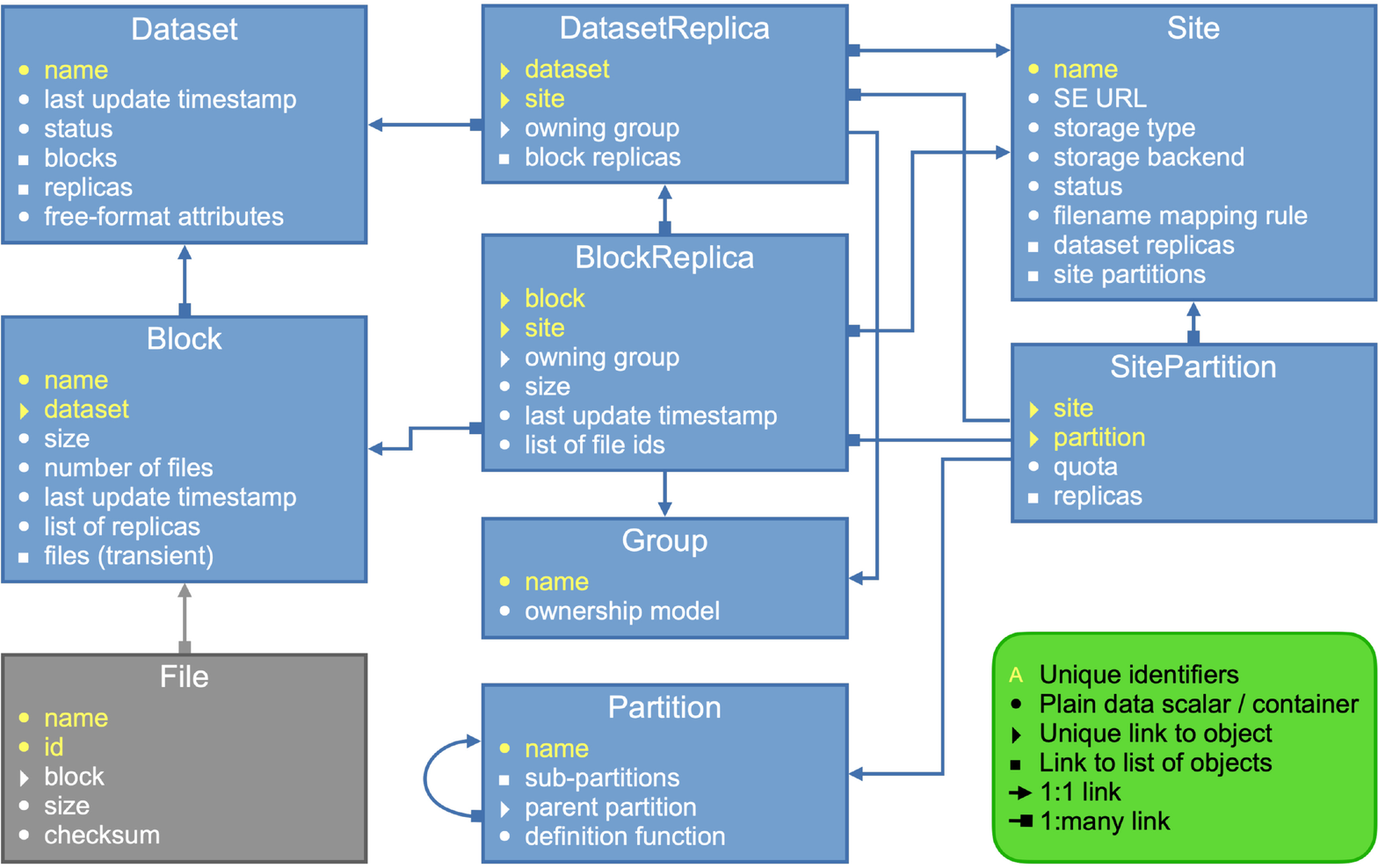}
  \caption{Dynamo objects in the inventory and their relations. Each box is a class representing an entity in the inventory, with examples of object attributes listed inside. Attributes labeled as unique identifiers are used to identify the object. The box for the File class is drawn with a different color from the others to indicate that the File objects are not in the inventory RAM image but are created transiently on demand. The links between the boxes specify the types of relations between the classes.}
  \label{fig:inventory}
\end{figure*}

The Dynamo system revolves around the inventory, and therefore the basic concepts of the system are best understood through the objects in the inventory. Figure~\ref{fig:inventory} lists the classes of objects in the inventory and their relations. We refer to the objects in Figure~\ref{fig:inventory} as Dynamo objects in the following.

In Dynamo, data are managed in a three-tiered hierarchy. At the bottom of the hierarchy is the \textit{file}, which naturally maps to a POSIX file but can also represent other types of data units. A file in Dynamo is the atomic unit of data transfer and deletion. The system has knowledge only of whether a file exists fully in a given storage unit or not; there is no concept of any intermediate states such as partially transferred files.

Files that share semantic properties are grouped into a \textit{dataset}, which is the highest level of the hierarchy. For example, experimental data from a continuous period in a year and a Monte Carlo simulation sample for the same physics process are each organized as a dataset.

Because datasets can greatly vary in size, the intermediate grouping of \textit{blocks} is introduced to facilitate various data management tasks. Blocks are non-overlapping subdivisions of datasets, consisting of one or more files. There is no guideline for how blocks should be formed, but the intention is that they are purely logistical units that are semantically indistinguishable within a dataset. A block is the algorithmic atomic unit of data in Dynamo. In other words, decisions to replicate, move, and delete data are taken on the level of either datasets or blocks, but not files. Therefore, the typical volume and number of files of blocks affect the balance between fine-grain control of data placement and management efficiency. In the CMS experiment, the concept of blocks is actually incorporated into the file catalogue, which is populated by the production systems. The typical block size in this case is thus driven by the data organization needs at the production stage. The size of a CMS dataset is anywhere between a few gigabytes to a few hundred terabytes, and a typical block of a large dataset contains 5 to 10 files, adding up to 10 to 20 gigabytes in volume.

Computing clusters and other storage elements across the globe are represented as \textit{sites} in Dynamo. Sites are only defined by their network endpoints for data transfer and deletion. Attributes such as the external network bandwidth, total storage capacity, and the number of associated compute cores that may utilize the data in the storage can be optionally assigned to sites.

A copy of a dataset or a block at a site is called a dataset or block \textit{replica}. Following the hierarchy between datasets and blocks, a dataset replica at a site consists of replicas of the blocks of the dataset at the site. A block replica is considered complete if copies of all constituent files are at the site, and incomplete otherwise. Similarly, a dataset replica is incomplete if any of the constituent block replicas are incomplete. A dataset replica with no incomplete block replica is complete if all blocks of the dataset have a copy at the site, and partial if replicas of only a subset of the blocks exist.

A \textit{partition} of the entire global storage system is a group of block replicas defined by a set of rules. For example, a partition can be defined by replicas of blocks belonging to datasets with specific name patterns. Partitions do not have to be mutually exclusive. Sites may set quotas for different partitions at their storage elements. Quotas are however not enforced by the Dynamo system core, and it is up to the individual Dynamo applications to decide to respect them.

Dynamo has a simple language set that consists of short human-readable predicates regarding datasets, blocks, their replicas, and sites. The predicates may refer directly to attributes of the objects such as their last update timestamps, or can involve dynamically computed quantities such as the total number of replicas that currently exist in the overall system. The language set is called the \textit{policy language} because its primary use is in setting data placement policies for the applications, but is available for any other part of the program. For example, the rules on block replicas defining the partitions are written in the policy language.

One of the attributes of a dataset or block replica is its owning \textit{group}. Ownership is an easy way to flag the use purpose of a data element. For example, in the CMS experiment, data managed by Dynamo are mostly used either for physics analysis or for production of derived-format data, with significantly different usage patterns. Therefore, block replicas are owned by analysis or production groups, and partitions and data management policies are set separately for the two ownership groups. Ownership is transferred under special circumstances, such as when a block replica owned by one group is scheduled simultaneously for deletion and replication under another owning group. Note that the block replica ownership is purely a logical concept within the Dynamo software and does not relate to file ownerships of managed data at the site storage elements.

\section{Details of the system components}
\label{sec:components}

\subsection{Dynamo server and the inventory}
\label{subsec:inventory}

The main function of the Dynamo server is to manage the inventory and to launch the applications. The inventory is constructed in memory during the startup phase of the Dynamo server and kept until the server process is terminated. The server process runs as a daemon in a loop of checking for new application to run, spawning an application child process if there is one, checking for inventory updates sent by write-enabled applications, and collecting completed applications.

The inventory object consists of simple Python dictionaries for datasets, sites, groups, and partitions, with the names of objects as the key and the objects themselves as values. The objects are interlinked to reconstruct their conceptual relationships. For example, a dataset object has a list of its replicas and a list of its constituent blocks as attributes, and the dataset replica and block objects each point back to the dataset object also as their associated dataset. See Figure~\ref{fig:inventory} for the full schematics of the relationships among the objects.

%To keep the memory footprint of the inventory manageable, file-level information is not kept in memory but is loaded from the persistence layer only when it is needed, such as when scheduling file transfers of a block, and is discarded immediately after use.

As mentioned in Section~\ref{subsec:system_design}, the inventory can be updated by write-enabled child processes (applications and web modules). Write-enabled processes commit the changes they made to the inventory at the end of execution by passing a list of updated and deleted Dynamo objects to the server process. The objects are serialized into respective representation text strings and sent through an inter-process pipeline. For updated objects, the server process deserializes the received objects and realizes their new states in the inventory, i.e., an object is created if it did not exist previously, and its attributes are updated otherwise. Similarly, for deleted objects, the server finds the corresponding object in the inventory using the unique identifier of the object and unlinks it. These changes to the inventory are optionally persisted immediately, if a persistence provider is configured accordingly.
% An example code snippet is provided below
%for a write-enabled application that sets the status attribute of datasets with no replicas in the
%system.

New applications are not started during the update, but the ones that have been already running at the start of the inventory update keep running with the pre-update inventory image. The web server is restarted upon completion of the update.
%
%\begin{lstlisting}
%# Make the inventory object available
%from dynamo.core.executable import inventory
%from dynamo.dataformat import Dataset
%
%# Loop over all known datasets
%for dataset in inventory.datasets.itervalues():
%
%    if len(dataset.replicas) == 0:
%        # Invalidate datasets with no replicas
%        print "Invalidating " + dataset.name
%        dataset.status = Dataset.STAT_INVALID
%        inventory.register_update(dataset)
%
%    else:
%        # Inspect the block replicas of incomplete dataset replicas
%        for replica in dataset.replicas:
%            if not replica.is_complete():
%                for block_replica in replica.block_replicas:
%                    print block_replica.is_complete(), block_replica.num_files, '/', block_replica.block.num_files
%\end{lstlisting}

\subsection{Applications, scheduler, and interactive sessions}

Actual data management tasks are performed by Dynamo applications, using the Dynamo server and the inventory as infrastructure. Dynamo application executables are single-file python scripts that are submitted to the server and executed asynchronously. Any valid python script will be accepted as an application. Submission is done through a TLS socket connection to a designated port the Dynamo server listens to, using a command-line client program called \texttt{dynamo}, included in the Dynamo package. The python script is sent over the network or, if submitted from the machine the server is running on, copied from a local path. Submitter of the application is authenticated with their X.509~\cite{x509} certificate. The certificate Distinguished Name must be authorized beforehand to run applications on the server. Once the submitter is authenticated and passes the authorization check, the application execution request is queued in the server and is picked up in one of the server loop iterations.

The application programs access the inventory object by importing it to the namespace through a statement like
\begin{verbatim}
from dynamo.core import inventory
\end{verbatim}
. Because the inventory object seen by a Dynamo application exists in the address space of the application process, contents of the inventory can be modified in any way within the program without affecting the server or other concurrently running applications. Write-enabled applications call the \texttt{update} and \texttt{delete} methods of the inventory object to register updated and deleted Dynamo objects, respectively, to be sent to the server process.

In a production environment, the application scheduler would be used to execute same sets of applications in sequences repeatedly. Multiple sequences can be managed concurrently, allowing, for example, having one sequence that executes the transfer request processing with high frequency while scheduling a thorough consistency check of the global storage system once per week. To create a sequence managed by the scheduler, a sequence definition file is submitted to the Dynamo server using the \texttt{dynamo} command. The sequence definition file uses a simple syntax to specify the applications to run, the order of execution, idle time between the executions, exception handling (ignore exceptions and move on to the next application; repeat the failed application; or repeat the entire sequence), and how many times the sequence should be repeated. The application scheduler runs in an independent thread within the Dynamo server process, cycling over the sequences and the applications therein indefinitely until the server process is stopped.

System administrators or authorized users who wish to explore the contents of the inventory interactively can also start an interactive session over the socket connection using the \texttt{dynamo} command. These sessions also run as child processes of the server and therefore have a fully constructed inventory object available as a python object. The interface for the interactive session resembles the prompt of the interactive mode of the Python interpreter. This feature is also useful for application developers for prototyping applications at a small scale.

\subsection{Web server}

The Dynamo web server is an optional child process of the Dynamo server. It communicates via FastCGI with an external HTTP(S) frontend server, which handles the HTTP requests and their TLS authentications. The web server first parses the requested URL of the incoming HTTP request passed from the frontend. The URL specifies whether a web page or a data service (REST API) is requested, and also the name of the module that provides the contents. If the module is with restricted access, the request must have come over HTTPS, and the Distinguished Name of the user certificate is checked for authorization.

The identified module is then called with the full detail of the HTTP request, including the query string contained in the URL or posted in the HTTP request body. The module returns an HTML document string or a Python dictionary depending on whether a web page or a data service is requested. The web server formats the returned value from the module into the final string passed back to the HTTP frontend, to be sent to the requesting user.

The list of modules, and therefore available web services, is easily extendable. Modules are written as Python classes with certain methods. The author of the module only needs to provide a mapping from the module name in the URL to the class, which can be picked up by the web server without stopping the Dynamo server.

A child process of the web server is spawned for each HTTP request. While a thread-based web server is more efficient in terms of resource usage than one that spawns a process for each request, process-based web server isolates each module in its own address space in the same way that Dynamo applications are isolated. This guarantees that modules with read-only access to the inventory do not modify the inventory image held by the Dynamo server, and intermediate updates by write-enabled modules do not interfere with other web modules and applications.

When write-enabled web modules, such as the API for data injection, are executed, updates are communicated to the Dynamo server in terms of Dynamo objects, through the same pipeline used by write-enabled applications, at the end of the module execution. The web server process then restarts itself to reflect the change in the inventory in the server process.

\subsection{Transfer and delete operations}
\label{subsec:transfer_and_delete_operations}

The interfaces to transfer and delete operations, at the block replica level and at the file level, are provided as Python modules within the Dynamo package. The two levels are separated so that applications can work at different degrees of abstraction. For example, an application that determines which dataset replicas should be deleted from a site can use the replica-level interface and thus does not need to consider any file-level information.

These Python modules merely define the programming interfaces for the transfer and delete operations. Specific plugins implement the actual functions, and it is up to the applications to configure the interfaces with appropriate plugins. Default plugins are included in the Dynamo package for both operation levels. The plugin for the replica-level operations writes the operation information to the registry database, to be read and acted on by the file operation manager (See Section~\ref{subsec:fom}). Two plugins are provided for the file-level operations. One communicates with an FTS3 server to issue and track file transfers and deletions. In case there is no FTS3 instance available, the other plugin can be used together with a standalone lightweight daemon (\texttt{dynamo-fileopd}), based on the GFAL2 library~\cite{gfal2}, also included in the Dynamo package. This daemon however is written to run on a single server and therefore is unable to manage a large number of parallel operations.

There are only a few functions that these plugins must implement, making it straightforward for an experiment with an existing data management tool to adopt Dynamo as its higher-level layer. As long as the existing tool exposes its transfer and deletion commands in an API, either at the level of files or some larger unit that corresponds to blocks, a plugin can be written and Dynamo can function completely agnostic to how the operations are performed.

\subsection{Parallelization}
\label{subsec:parallelization}

Multiple Dynamo servers, each equipped with its own inventory, can be linked into a single server cluster for load balancing and high availability. The cluster is formed sequentially; when the first two servers are connected, one of the nodes assumes the role of the master server, which holds the messaging board (a set of database tables which can be connected remotely), through which the members of the cluster communicate. Subsequent servers join the cluster by first connecting to any of the nodes in the cluster. Upon establishing the connection, the node that was already in the cluster will send the host name of the master server to the new node, allowing the latter to connect to the correct master server if it is different from the former.

Heart beat signals are sent from the member nodes of the cluster to the master node at a regular interval, and as a response, the contents of the messaging board are sent back. Thus, although the master server acts as the communication hub of the members of the cluster, all nodes are on equal footing in terms of their information content. This ensures that there is no single point of failure in the cluster. If a non-master server fails and does not send a heart beat for some time, it is trivially excluded from the cluster. If the master server fails, the second server registered in the messaging board becomes the new master server.

When a new node connects to a cluster, its inventory content is copied from one of the nodes already in the cluster. The inventories in the cluster are kept in synchronization by allowing only one write-enabled application or web module to run at a time throughout the cluster, using a locking mechanism provided through the messaging board. The updates made to one inventory are then broadcast to the linked servers before the lock is released. The same serialized queue of Dynamo objects used to update the local inventory is used for remote updates.

To create a load-balancing cluster where multiple nodes are accessed under a single host name in e.g. a round-robin mechanism, services such as \texttt{keepalived} must be run on top of the Dynamo cluster. Dynamo itself only provides the machinery to operate parallel linked server instances.

\section{Applications}
\label{sec:applications}

While Dynamo server manages the inventory image, it is the individual applications that utilize the information in the inventory and carry out the actual data management tasks. As noted in Section~\ref{subsec:system_design}, any valid Python program can become a Dynamo application, allowing the user of the system to define and execute arbitrary new tasks under the system.

\begin{figure*}
  \includegraphics[width=0.975\textwidth]{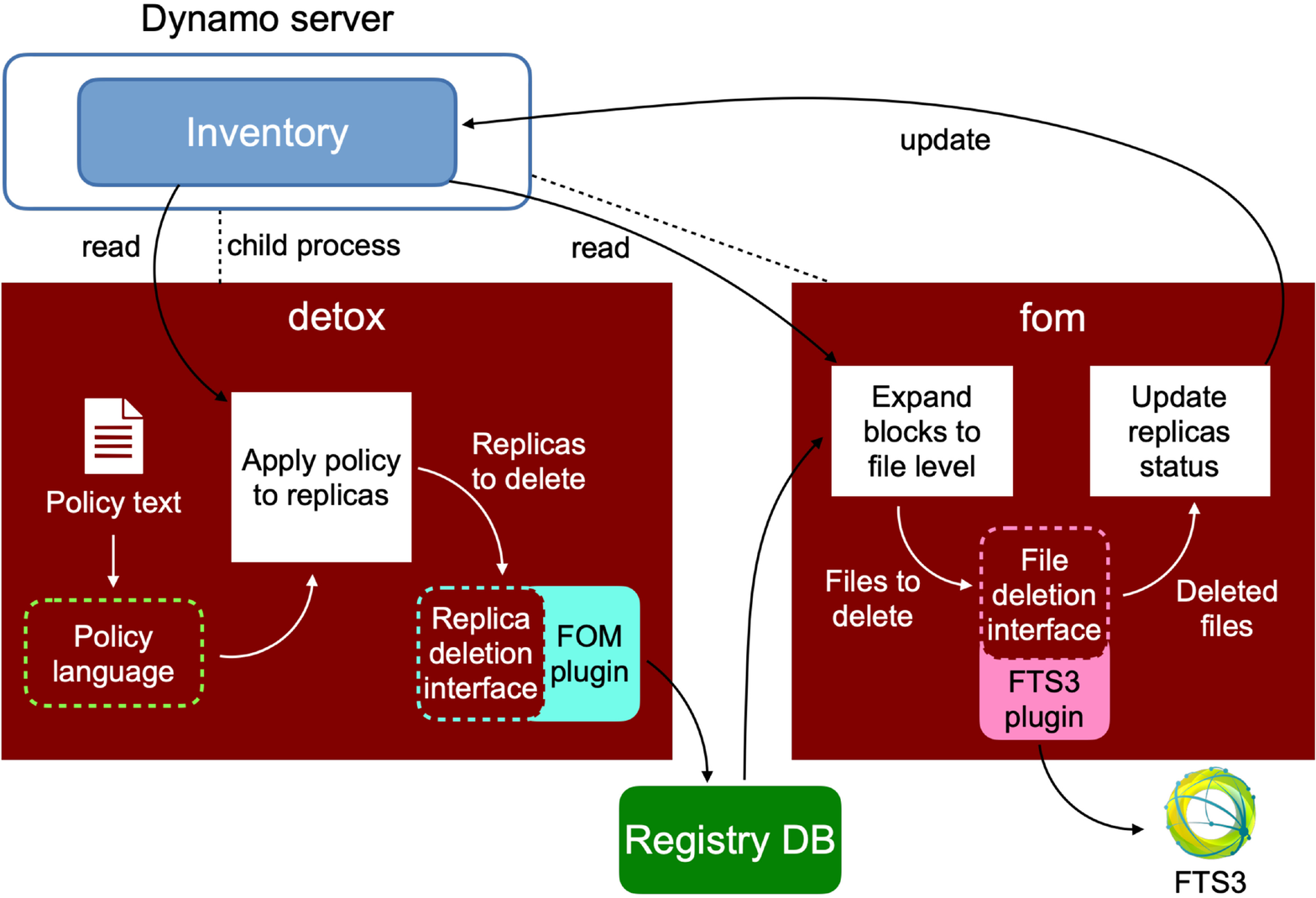}
  \caption{Schematic of the data deletion application \texttt{dynamo} and the file operation manager (\texttt{fom}). Solid lines represent the flow of information. Boxes with dashed outline are utility components of the Dynamo package, i.e., software provided in the package but are not parts of the system core (Dynamo server). As explained in Section~\ref{subsec:system_design}, the registry database facilitates asynchronous inter-process communications.}
  \label{fig:detox_diagram}
\end{figure*}

As an example, Figure~\ref{fig:detox_diagram} shows a schematic of how two default applications, \texttt{detox} and \texttt{fom}, work in tandem. As Dynamo applications, both are child processes of the Dynamo server process, and read and possibly update the inventory. Utility components of the Dynamo software package, such as the policy language and the replica or file deletion interfaces, are used within the applications. The deletion interfaces are configured with backend plugins that carry out the deletion operations.

This section describes these and other default applications for common tasks that a data management system would perform. The source code for these applications is included in the standard Dynamo software package.

\subsection{Data deletion: \texttt{detox}}

The dynamic management of space adheres to two fundamental principles: firstly, the utilization should not go too close to $100\%$ of the available disk space for reasons of flexibility and stability; secondly, having a substantial fraction of empty, but in principle available space resources is not an economic approach. A proper, high utilization is therefore desired.

A deletion agent application, called \texttt{detox}, is run regularly to prevent storage sites from overflowing. The application evaluates a policy at run time to determine if deletions are necessary and allowed at a given site. A policy in \texttt{detox} consists of general directives, such as which sites and partitions are cleaned up, and a list of rules and actions that are taken on replicas satisfying the rules. In a standard \texttt{detox} operation, dataset replicas at sites with occupancy above the upper watermark are marked for deletion in the order specified in the policy until their projected occupancy after deletions have been brought down to the lower watermark.

Data attributes, which are freely configurable in the \texttt{detox} libraries, are evaluated and matched line by line to the rules in the policy. These rules make use of data attributes like the popularity of a dataset or whether it has a replica on tape storage. The attributes are set by dedicated producers at run time. Data are sorted into cannot-be-deleted, can-be-deleted, or must-be-deleted categories. Data in the can-be-deleted category are deleted if the site requires cleanup (e.g. because the occupancy is above the upper watermark).
%\begin{figure}[h]
%\centering
%\includegraphics[width=0.975\textwidth]{figures/policy.pdf}
%\caption{\label{fig:policy}Understanding the policy syntax}
%\end{figure}
%Figure~\ref{fig:policy} shows an example for such a policy. Data attributes (in this case \texttt{dataset.status}) are evaluated at runtime. 

The following is an example of a \texttt{detox} policy. This policy applies to all sites and aims to keep the site occupancy between upper and lower watermarks of 90\% and 85\%, respectively. Lines starting with action statements \verb|Delete|, \verb|Protect|, and \verb|Dismiss| are applied to dataset replicas on a first-match basis. In the example, replicas of invalidated datasets are deleted unconditionally. For valid datasets, replicas are protected if they do not have full copies on tape archives, but otherwise can be deleted if their usage rank (which roughly corresponds to the number of days since the last usage) is greater than 200. Replicas that do not match any of the above criteria are protected by the last \verb|Protect| line, which sets the default action.

\begin{footnotesize}
\begin{verbatim}
    On site.name in [*]
    When site.occupancy > 0.9
    Until site.occupancy < 0.85
    Delete dataset.status == INVALID
    Protect dataset.on_tape != FULL
    Dismiss dataset.usage_rank > 200
    Protect
    Order decreasing dataset.usage_rank \
      increasing replica.size
\end{verbatim}
\end{footnotesize}

Custom locks preventing items from being deleted from sites can be placed from the REST API of the Dynamo web server. A web module accepts lock requests from authorized users and records the specifications of the dataset and block replicas to be locked in the registry database. These locks will be respected by \texttt{detox} upon adding the line
\begin{footnotesize}
\begin{verbatim}
    ProtectBlock blockreplica.is_locked
\end{verbatim}
\end{footnotesize}
to the policy.

%\begin{lstlisting}[language=bash]
%  $ curl --cert /path/to/cert --key /path/to/key %'https://dynamo.mit.edu/data/detox/lock/lock?item=NameOfItem&expires=\
%        2019-06-30&site=T2_CH_CERN&comment=HeavyIon-Reprocessing'
%\end{lstlisting}

\begin{figure*}
\begin{centering}
\includegraphics{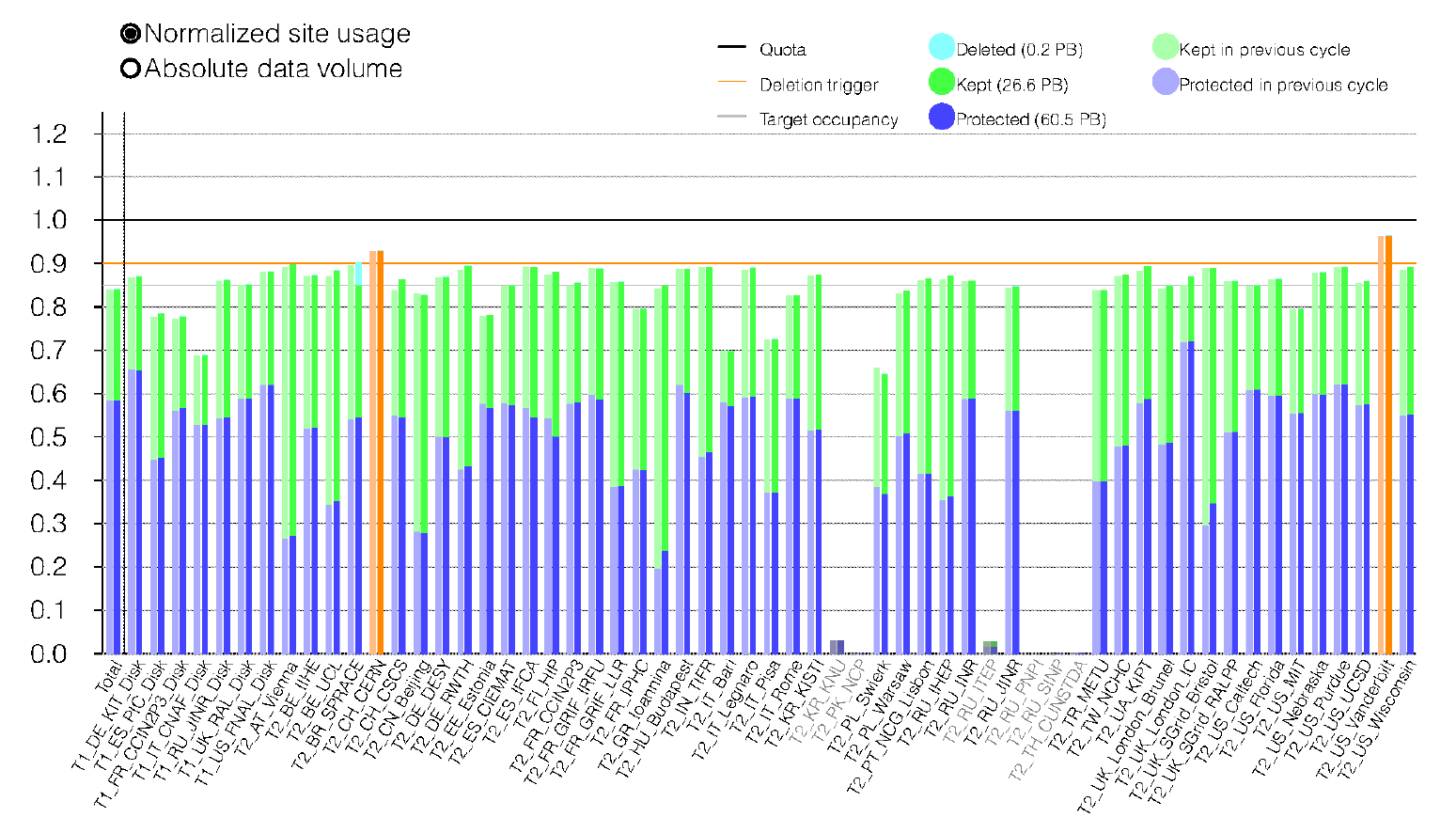}
\end{centering}
\caption{A snapshot of the site usage after a \texttt{detox} cycle has completed. Blue bars correspond to data that cannot be deleted because it is explicitly locked or protected for another reason. Green bars indicate the data for which there were no policy match and can therefore be deleted if the corresponding site occupancy exceeds the upper watermark at 90\%. Storage sites with orange bars have protected data above the upper watermark. Sites whose names are grayed out are currently dysfunctional and consequently have been emptied.}
\label{fig:detox_snapshot}
\end{figure*}

Figure~\ref{fig:detox_snapshot} shows a snapshot of the disk utilization of sites in the CMS experiment after a \texttt{detox} cycle has run. In this cycle, 0.2 petabytes of can-be-deleted data have been deleted because the occupancies of the respective storage sites were above the upper watermark. Snapshot plots like this are generated in the \texttt{detox} web page, included as a default web server module in the Dynamo package.

The \texttt{detox} application can be run in simulation mode to easily gauge the effect of a new policy on the system state without actually performing the deletions. Using this feature, \texttt{detox} is also being used in the CMS experiment to plan, organize, and execute dedicated deletion campaigns to remove obsolete datasets from tape archives on a yearly basis.

\subsection{Data replication: \texttt{dealer}}
\label{subsec:dealer}

Various reasons exist for why a specific piece of data should be replicated at specific sites or unspecifically across the global storage pool: a high demand by users; (temporary) unreliability of specific storage sites; desire to evenly distribute critical datasets to prevent imbalances and therefore single-points-of-failure in the system; recall from tape; initial data injection; etc.

An application called \texttt{dealer} is run in a regular cycle to evaluate the replication requests and determine the data copies to make. The application collects the requests from its various plugins, each representing a different reason for requiring data replications.

The different plugins are described briefly in the following.

\begin{itemize}

\item The \textit{popularity plugin} proposes replications of datasets that are frequently read by analysis users. The information on access frequency, provided by an external service, are combined with other factors, such as the size of the dataset, into a weight factor assigned to each dataset. It should be noted that this weight factor is a good place for the incorporation of machine learning algorithms, like reinforcement learning, to predict which datasets will be accessed in the near future and hence have them ready and available on multiple sites to facilitate their access for the users.

\item The \emph{balancer plugin} aims at replicating data present only at a single site (``last copy'') which has a large fraction of protected data. It will propose to replicate these data at a second destination, so that the protected space can be freed up at the original site and the protected data are evenly distributed across the storage sites. This minimizes the risk of data unavailability and creates a contingent of data at each site that can be deleted upon demand.

\item The \emph{enforcer plugin} deals with static rules for replication. It will try to accommodate special rules for data placement, such as ``The replication factor for datasets of type $X$ on continent $A$ should equal 2''. A special attribute on dataset and block replicas indicate that the replica is being managed by an enforcer rule. Detox policy should be set up to protect replicas with this attribute set to true.

\item When a storage site will be unavailable for an extended period of time, it is advised to remove all data from the site so that user jobs do not try to access data and get stuck or fail in the attempt of doing so. The \emph{undertaker plugin} proposes replication of dataset replicas that are unique at specified problematic sites. Replicas copied to other sites will then be deleted by \texttt{detox}, clearing out the problematic sites. Figure~\ref{fig:detox_snapshot} displays sites in non-functional state as greyed out and cleaned out.

\item The \emph{request plugin} works in combination with a REST API web module, to which production and analysis systems (and any authorized users) make explicit data replication requests. The production system would request data replications to prepare the input for its tasks or to consolidate the output of the production jobs that are scattered across the distributed computing system. The requests from the analysis system may occur when e.g. the dataset to be analyzed has replicas only in the tape archives.

\end{itemize}

The decision on which datasets to finally replicate is made from the proposed candidates at random, taking into account a configurable priority value assigned to the proposing plugin, until the target occupancy of the storage sites is met (also considering the projected volume of ongoing transfers) or until a certain threshold is reached which limits the amount of data replicated per \texttt{dealer} cycle.

\subsection{Site Consistency}

The application \texttt{dynamo-consistency} checks the consistency between Dynamo's inventory and files actually located at managed sites.  Even though Dynamo controls and tracks the history of file transfers and deletions at its sites, a separate check is needed to ensure that files are not lost or accumulated due to user or system errors.  Actual site storage content and the inventory can become inconsistent either when files that are supposed to be at a site according to the inventory are deleted or inaccessible (missing files) or when files that are not cataloged in the inventory exist (orphan files). Missing files cause failures of block transfer requests. Jobs that are assigned to run at the site with missing files, assuming to read these files locally will fail, or if there is a backup scenario will be inefficient as they are forced to read the files remotely instead. Orphan files on the other hand lead to wasted disk space. \texttt{dynamo-consistency} can be run regularly to check consistency by listing the contents of each remote site and comparing the results to the inventory.

Sites managed by Dynamo may all employ different mass storage technologies and their remote interfaces. \texttt{dynamo-consistency} supports remote site listing using XRootD Python bindings, \texttt{xrdfs} subshell, and the \texttt{gfal-ls} CLI of the GFAL2 library. The base lister class is easily extensible in Python, allowing for new site architectures to be checked by \texttt{dynamo-consistency}.

Files matching filtering criteria, which are configurable, are excluded from being listed as missing or orphan, even if they are inconsistent with the inventory. For example, a file with a recent modification time may appear as an orphan only because there is a time lag in updating the inventory, and thus should be exempt from listing. Also, certain paths can be excluded from the check via pattern matching. The consistency check also does not report files that are present in Dynamo's pending deletion and transfer requests in order to not trigger redundant actions.

% Original paragraph by Dan
%The default executable performs the check as expected, listing files that are not tracked by Dynamo
%as orphans and listing files that are not found at sites as missing, with the exception of a few
%configurable filters.  Dynamo Consistency avoids listing orphan files that have a modification time
%that is recent.  Paths to avoid deleting can also be set.  Deletion and transfer requests that are
%queued are also used to filter the final report to avoid redundant actions from Dynamo.

% I think we can skip this (Y.I.)
%In addition to tracking the consistency between Dynamo's inventory and physical site storage,
%Dynamo Consistency can report all remote files older than a certain age in general directories.
%These files can also be filtered with path patterns, just as the regular consistency check.
%The time-based only reporting allows for cleaning of directories that Dynamo does not track.
%This is a setting recommended for large file systems that are written to with a high frequency.

Summaries of check results, as well as the statuses of running checks, are displayed in a webpage. The page consists of a table that includes links to logs and lists of orphan and missing files. Cells are color coded to allow operators to quickly identify problematic sites. Historic summary data for each site is also accessible through this page.

%If the available configuration options and listers are not enough,
%advanced users can also directly use the Python API to run a custom consistency check.
%For more details on the Dynamo Consistency package, see \sphinxurl{https://dynamo-consistency.readthedocs.io}.

\subsection{File Operations: \texttt{fom}}
\label{subsec:fom}

The Dynamo software package contains an application for scheduling and monitoring file transfers and deletions named \texttt{fom}. As noted in Section~\ref{subsec:transfer_and_delete_operations}, the transfer and deletion operation backend is decoupled from the Dynamo core, allowing experiments with existing file operations programs to retain them by writing a simple plugin upon adopting Dynamo. When no such program exists or a full-stack standalone operation of Dynamo is desired, \texttt{fom} can be used as the file operations provider.

To use \texttt{fom}, applications must use the replica-level transfer and deletion operations interface, as described in Section~\ref{subsec:transfer_and_delete_operations}, configured with \texttt{fom} as the backend. Transfer and deletion decisions made by the applications are then written to the registry database, to be read later by \texttt{fom}.

Because \texttt{fom} is a Dynamo application, it cannot be run as a daemon, and therefore does not monitor the progress of file transfers and deletions continuously. In fact, \texttt{fom} must itself delegate the management of transfers and deletions to a backend daemon program. At each execution, \texttt{fom} translates block-level transfer and deletion commands in the registry database into file-level information and issues the corresponding commands to the file-level operations backend. It also queries the backend to collect the reports on operations started in previous execution iterations. The reports (success or failure) are then used to update the inventory. The backend daemon can either be FTS3, \texttt{dynamo-fileopd}, or any other service if a plugin can be written.

\begin{figure*}
\centering
\includegraphics[width=0.8\textwidth]{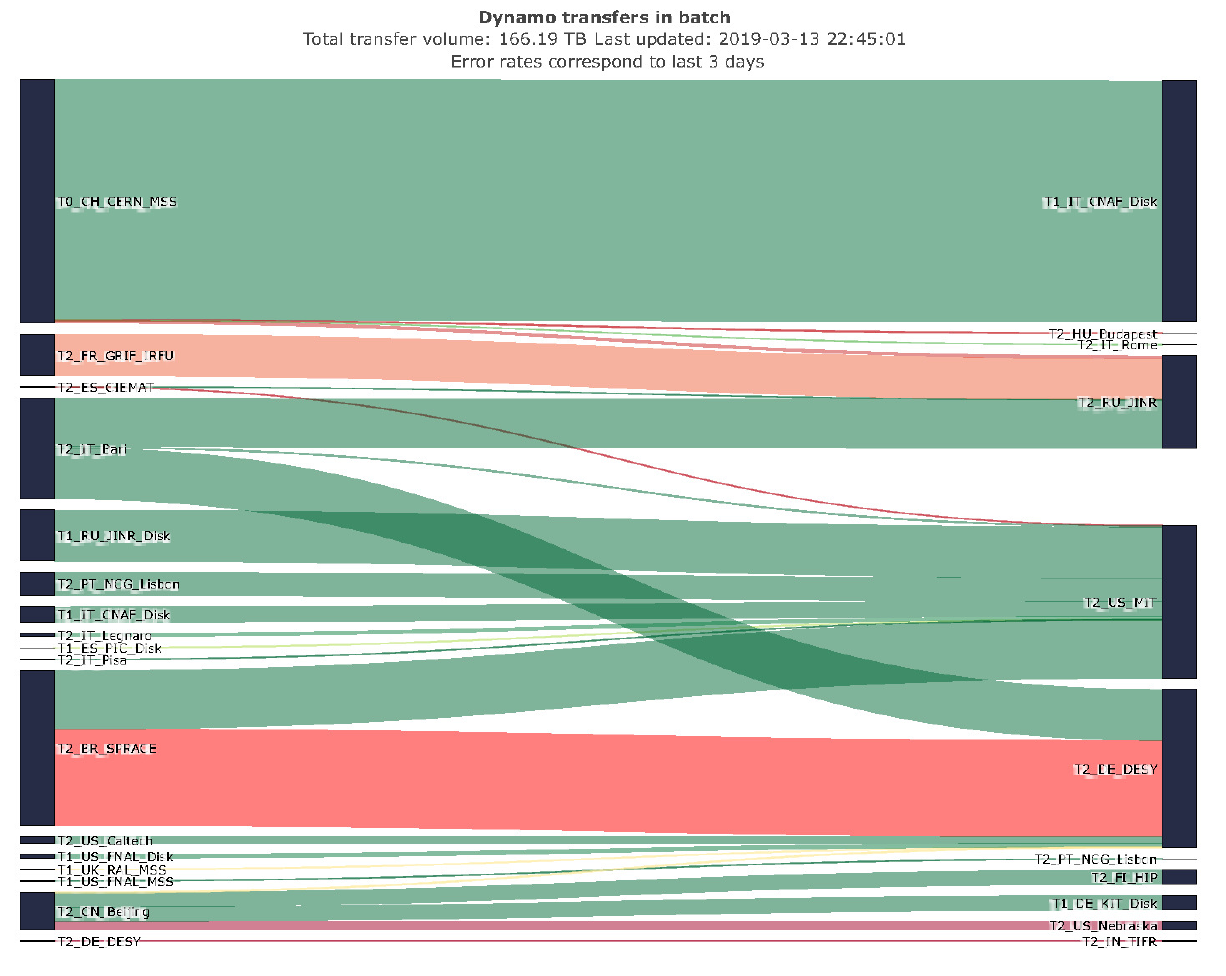}
\caption{The transfer flow diagram from the CMS Dynamo instance. The width of the bands represent the total volume of scheduled transfers between the sites indicated at the two ends. The color scale corresponds to the fraction of failed transfer attempts for a period of time (here: the last three days). The diagram helps to quickly identify problematic links that then can be investigated closer.}
\label{fig:transfers_sankey}
\end{figure*}

Transfer success and failure reports collected from the backend are also used to evaluate the quality of links between the sites. Figure~\ref{fig:transfers_sankey} is a plot showing ongoing transfers between different sites, where the widths of the bands represent the total volume of scheduled transfers and the colors of the bands encode the historical link quality information. This diagram is available in a web page generated by one of the default web server modules in the Dynamo package.

\begin{figure*}
\centering
\includegraphics[width=0.9\textwidth]{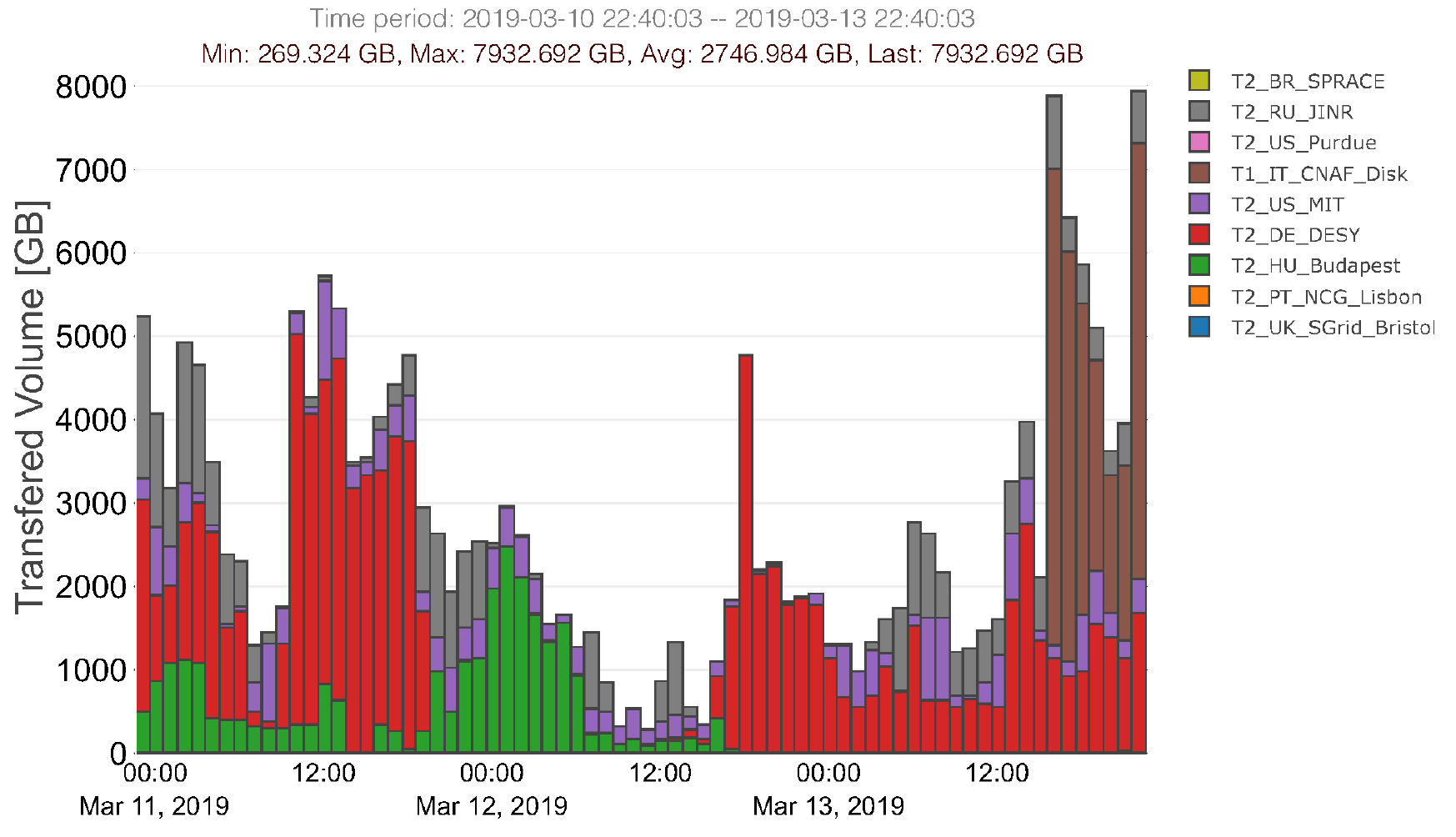}
\caption{History of transferred volume per hour, color-coded by transfer destinations.}
\label{fig:transfers_history}
\end{figure*}

Another web module exists to display the volume and rate of transfers as a time series. An example of the transfer volume history plot is in Figure~\ref{fig:transfers_history}.

\subsection{REST API}

Although they are not strictly Dynamo applications, Dynamo web server modules, and consequently the REST API, also run as child processes of the Dynamo server process with access to the inventory image. The REST API allows general users to access the information in the inventory through a number of remote calls described in this section. Because the inventory is fully loaded onto RAM except for the information on files, responses to most API calls do not involve database I/O and thus are fast.

There are two distinct types of API calls available. The first, \textit{writing} type invokes operations that modify the state of the inventory, such as transfer and deletion of dataset and block replicas, or injection of new datasets and blocks. Only authorized users are allowed to execute these calls. These calls are exclusive, i.e., when a writing call is made while another writing call is being processed, the second call fails and reports HTTP error code 503 (Service Unavailable). Failure is necessary since the web server process is restarted to refresh the inventory image as soon as the first writing call completes, which implies that existing connections made to the server would have to be closed. Writing calls also fail during the execution of write-enabled Dynamo applications. The second, \textit{read-only} type of API calls allows general users to obtain various information about the inventory without changing its state. These calls do not have authorization restrictions and can be executed in parallel with any other web modules or applications running concurrently.
 
The list of existing REST API URLs can be found in Appendix~\ref{appendix:listOfApis}.
% The names, input, and output parameters are chosen in such way that would allow ongoing CMS operations to start using Dynamo APIs instead by simply switching URL address to the one reserved for Dynamo.

All incoming HTTP requests are sorted into two separate queues for further analysis for possible development of the API. The first queue contains calls that are malformed or do not exist at the moment. In this way users can signal the developers what they would like to have available in the future. The second queue contains valid calls to existing functions. Analysis of the second queue can shed light on which calls are popular and which ones can be possibly made obsolete.

The Dynamo web server has two layers of defense against distributed denial of service (DDoS) attacks. First layer is a \textit{DenyHosts}~\cite{denyhosts} service that blocks well-recognized sources of attacks. The second layer analyzes the HTTP request queues. If the frequency of correct or malformed requests from a single source passes a certain level that is deemed intrusive, the issuing address is automatically blacklisted in the firewall to prevent any further connection.
%We have to set up a simulation of DDos attack in order to test the system and determine at what frequency dynamo needs to act.

\section{Use Cases}
\label{sec:usecases}

\subsection{CMS experiment}

Dynamo has been in use by the CMS collaboration since the beginning of the LHC Run 2. This CMS instance handles several hundreds of petabytes of recorded and simulated experimental data stored across a worldwide computing grid, and has proven to work well at these scales and volumes. There are some noteworthy points from the operational experience.

First, loading the inventory at the startup phase of the Dynamo server is not instantaneous for a system of this scale, but completes within a manageable time. The CMS experiment has roughly $5\times 10^5$ datasets, $5 \times 10^6$ blocks, $10^6$ dataset replicas, and $10^7$ block replicas, and the inventory construction takes approximately 15 minutes using a machine with an Intel\textregistered Xeon\textregistered Gold 6134 CPU and MariaDB\cite{mariadb} database on a solid-state drive for persistence. The constructed inventory has a size of approximately 8 gigabytes.

Construction of the inventory for the CMS experiment would require a substantial amount of time, if done from scratch. With the order of five machines running parallel Dynamo servers, there is very little risk of losing the information in the inventory. However, even in the case of a catastrophic failure, Dynamo can be started with no block and dataset replicas registered in the inventory, and \texttt{dynamo-consistency} can be used to detect which files, and thus block and dataset replicas, are at each site. Since listing the content of one of the largest CMS site with 20 petabytes of disk storage with \texttt{dynamo-consistency} (remotely) takes roughly 50 hours, such recovery procedure (running many \texttt{dynamo-consistency} application instances in parallel) would take a few days.
%Locally produced inventories of files though could be produced in less than an hour and fed to the consistency agent, but would require manual intervention.

Applications also do not execute instantaneously in the CMS instance, but complete within practically reasonable time. For example, it takes at least 15 minutes to complete a full cycle of routine \texttt{detox} execution, in which the occupancy of in the order of 60 sites are checked and the dataset replicas to delete are determined. The execution time is driven by both the number of dataset replicas to consider and the number of policy lines, which is 33 at the time of writing, to evaluate for each replica. Similarly, a routine \texttt{dealer} cycle evaluating replication requests from all of the plugins listed in Section~\ref{subsec:dealer} takes 10 minutes. Because the datasets in this instance are typically accessed by non-interactive batch jobs, execution time scale of less than $\mathcal{O}(1)$ hour is acceptable.

\begin{figure*}[htbp]
\centering
\includegraphics[width=0.65\textwidth]{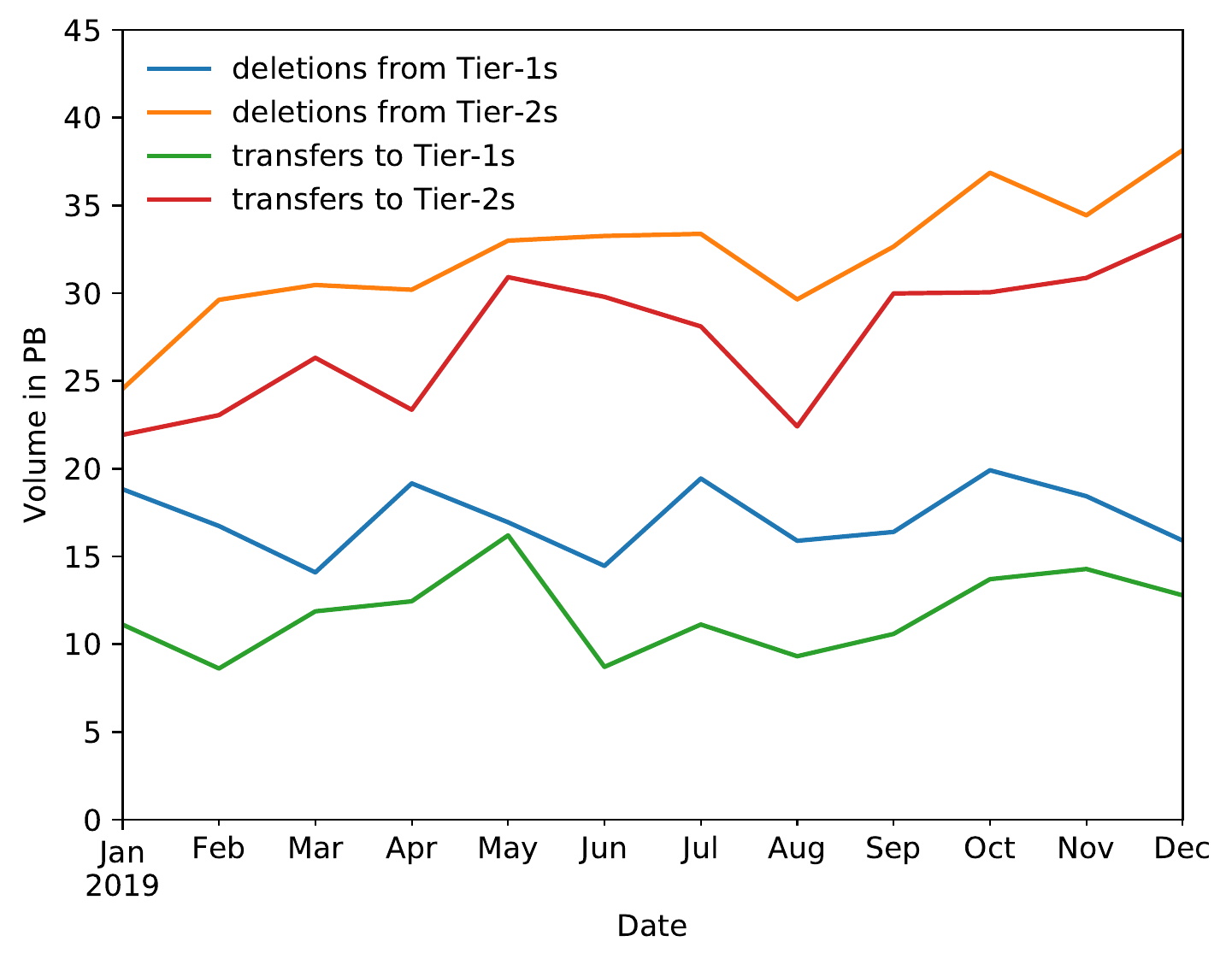}
\caption{Monthly data volume transferred to and deleted from the Tier-1 and Tier-2 disk sites in CMS in 2019.}
\label{fig:yearly_volume}
\end{figure*}

Figure~\ref{fig:yearly_volume} demonstrates that the CMS instance of Dynamo is able to operate stably at the required scale. The figure shows the monthly total of data volume transferred to and deleted from the CMS Tier-1 and Tier-2 sites by Dynamo for the year 2019. Several dozens of petabytes were moved and deleted per month. Here, deleted datasets are typically the unpopular ones, perhaps because of their age, and the transfers replaced them with high-demand datasets. Thus Dynamo creates a ``data metabolism'' of the CMS experiment to utilize the limited disk space most effectively. There are more deletions than transfers because the simulation datasets are constantly being generated at the sites, acting effectively as sourceless transfers.

\begin{figure*}[htbp]
\centering
\includegraphics[width=0.53\textwidth, height=0.6\textwidth]{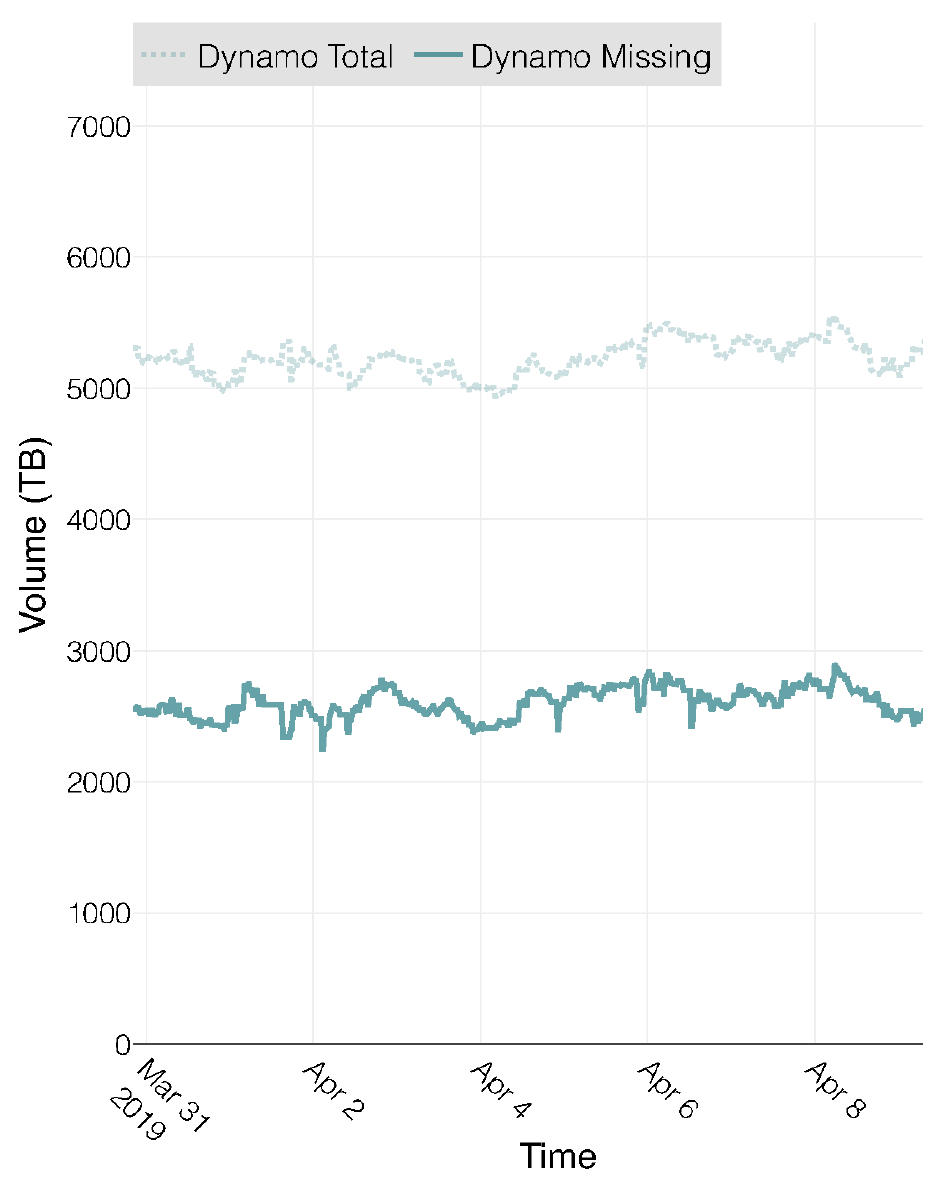}
\caption{Weekly overview of the missing and total volume of incomplete transfers. The missing volume roughly follows the total volume of transfers.
  %The backlog corresponds to the equilibrium reached between the data transfers assigned per cycle and the performance of the transfer system governed by link quality, storage site reliability, etc.
}
\label{fig:dealer_tracking}
\end{figure*}

While file deletion operations usually complete quickly, transferring terabyte-size datasets can take from several hours to even several days depending on the connectivity and available bandwidth between the involved sites. Therefore, at any given moment, there is a queue of incomplete transfers in the CMS Dynamo instance. The \texttt{dealer} application has a feedback mechanism that suppresses new replications when the queue of pending transfers is too long, but if this mechanism is invoked too frequently, the system will be slow to respond to e.g. a surge of popularity of certain datasets. Therefore, a limit must be placed on the total volume of dataset replications to be requested in a single \texttt{dealer} execution to ensure a healthy data metabolism. Experience has found that ordering at most 200 terabytes worth of replicas per \texttt{dealer} execution iteration, repeated after roughly one hour of interval, allows creation of sufficient amount of new replication orders at each cycle while keeping the utilization of the transfer system high. Figure~\ref{fig:dealer_tracking} shows a time series of the total volume of data replication (``Total'') scheduled by \texttt{dealer} and its subset that has not completed yet (``Missing''). As individual dataset replications progress, the missing volume are brought lower, and when the replication of a dataset completes, its volume is taken out of the total. In the figure, the ``Total'' curve stays at a similar level because new replication requests are constantly being made, and the ``Missing'' curve follows the ``Total'' curve because the overall CMS storage system is able to handle this scale of transfers.
%Different systems performing differently well in terms of average network speed and link quality between storage sites will exhibit a different backlog of data to be transferred, but an equilibrium between the transferred and newly subscribed volume will be reached regardless, see Fig.~\ref{fig:dealer}.

\begin{figure*}[htbp]
\centering
\includegraphics[width=0.975\textwidth]{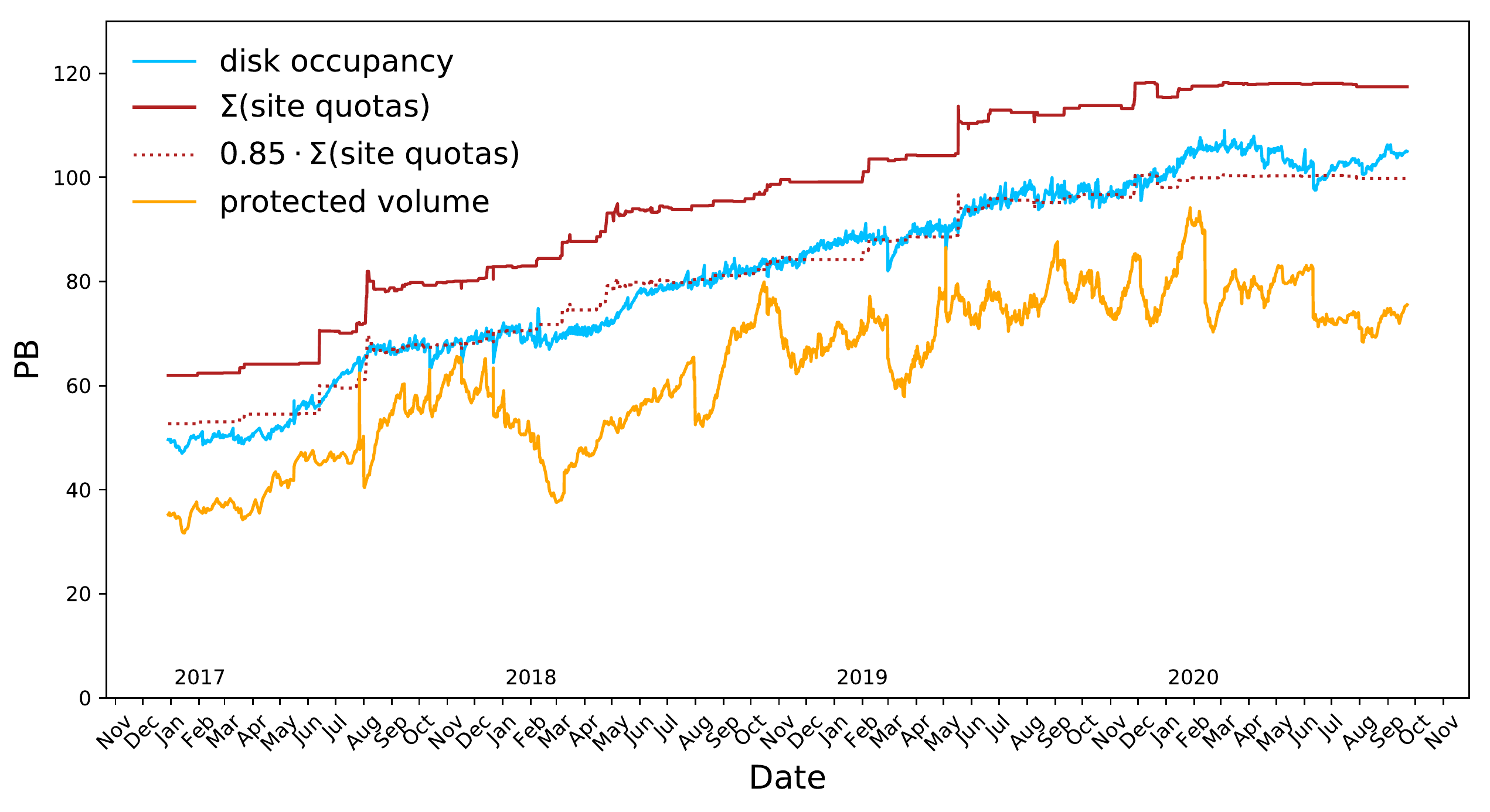}
\caption{Time series of the disk space managed by Dynamo. The occupied volume (blue line) is held at a target size of 85\% of the quota pledged by the storage sites (red dotted and solid lines, respectively). An orange line indicates the volume per Detox cycle that was protected according to the policies and could not be deleted.
}
\label{fig:occupancy_history}
\end{figure*}

As seen from Fig.~\ref{fig:occupancy_history}, Dynamo is able to hold the data volume on disk at around 85\% of the total space pledged by the storage sites for the entirety of 2017--2020. A target occupancy of 85\% is an input parameter to Dynamo and is chosen as a good compromise between a high disk utilization and the availability of free buffer space, operating the system sufficiently far away from the critical scenario of all disks running full.

\subsection{Local university research group}

To evaluate the behavior of Dynamo in a different scenario, a full-stack instance is installed at a local university research group. This instance manages two storage sites, where one site is the ``master'' storage that holds all of the approximately 600 terabytes of data under management, and the other site, with a smaller capacity of 150 terabytes, is the cache storage for locally running jobs. Thus, the primary purpose of Dynamo in this instance is to keep the cache storage filled with datasets that are the most useful for the ongoing analyses at any given moment.

Managed data in this instance are organized into approximately $4 \times 10^3$ datasets, with dataset sizes varying from a few gigabytes to a few tens of terabytes. There are $9 \times 10^4$ blocks and $5 \times 10^5$ files, with a typical file size of 2 gigabytes. At this scale, server startup (loading inventory) completes in 20 seconds, and the execution of \texttt{detox} and \texttt{dealer} only takes a few seconds. This enables, in particular, running \texttt{dealer} every minute or more frequently, enabling a virtually real-time Dynamo response to user demands.

\section{Summary}

A data management software named Dynamo was created to satisfy the operational needs of the CMS experiment. Dynamo consists of a main server, which holds the image of the managed storage system in memory, and several applications, which perform the actual data management tasks. Its extensive web interface allows remote users and external services to monitor the status of various operations and to interact with the system. While the system was designed with usage in the CMS experiment in mind, its architecture easily accommodates different use cases at a wide range of installation scales.

\section{Software availability}

Dynamo standard software package is available at \\
\verb|https://github.com/SmartDataProjects/dynamo|. \\
The \texttt{dynamo-consistency} application is available at \verb|https://github.com/SmartDataProjects/| \\ \verb|dynamo-consistency|.

\section{Conflict of Interest}
On behalf of all authors, the corresponding author states that there is no conflict of interest.

\begin{acknowledgements}

This material is based upon work supported by the U.S. National Science Foundation under Award Number PHY-1624356 and the U.S. Department of Energy Office of Science Office of Nuclear Physics under Award Number DE-SC0011939.

Disclaimer: ``This report was prepared as an account of work sponsored by an agency of the United
States Government. Neither the United States Government nor any agency thereof, nor any of their
employees, makes any warranty, express or implied, or assumes any legal liability or responsibility
for the accuracy, completeness, or usefulness of any information, apparatus, product, or process
disclosed, or represents that its use would not infringe privately owned rights. Reference herein to
any specific commercial product, process, or service by trade name, trademark, manufacturer, or
otherwise does not necessarily constitute or imply its endorsement, recommendation, or favoring by
the United States Government or any agency thereof. The views and opinions of authors expressed
herein do not necessarily state or reflect those of the United States Government or any agency thereof.''

The authors thank the CMS collaboration for extensive feedback and support.

\end{acknowledgements}

\begin{appendices}
\section{List of APIs}
\label{appendix:listOfApis}

\noindent\emph{Groups:} A List of known groups.\\
Input options are:
\begin{itemize}
    \item \texttt{required}: none
    \item \texttt{optional}: group (name of the group)
\end{itemize}
Output:
\begin{itemize}
    \item \texttt{name}: group name
    \item \texttt{id}: group id
\end{itemize}
\ \\
\emph{Nodes:} A list of sites known to Dynamo.\\
Input options are:
\begin{itemize}
    \item \texttt{node}: Dynamo site list to filter on~(*)\footnote{ (*) means any sequence of chracters. For example, `T2\_US*' would match any site that starts with `T2\_US'}
    \item \texttt{noempty}: filter out sites that do not host any data
\end{itemize}
Output:
\begin{itemize}
    \item \texttt{name}: Dynamo site name
    \item \texttt{se}: node type, can be `Disk' or `MSS' (i.e., tape)
    \item \texttt{id}: unique site id assigned intrinsically by Dynamo
\end{itemize}
\ \\
\emph{Datasets:} Basic information about datasets.\\
Input options are:
\begin{itemize}
    \item \texttt{dataset}: dataset name, can be multiple (*)
\end{itemize}
Output:
\begin{itemize}
    \item \texttt{name}: List of the matched datasets that include full dataset name, size, number of file, status, and type.
\end{itemize}
\ \\
\emph{Subscriptions:} Show current subscriptions (dataset and block replicas) and their parameters.\\
Input options are:
\begin{itemize}
    \item \texttt{dataset}: Dataset name~(*)
    \item \texttt{block}: Block name~(*)
    \item \texttt{node}: Site name~(*)
    \item \texttt{group}: Group name
    \item \texttt{custodial}: y or n, indicates if it assigned to tape storage
\end{itemize}
Output:
\begin{itemize}
    \item \texttt{dataset}: List of datasets, each list item contains a dataset replica if a complete replica exists, and a list of blocks replicas if not
    \item \texttt{block}: Each item in list of blocks contains a block replica
    \item \texttt{subscription}: contains node (site name), id (site name), request (request id), node\_files (number of files at this site), node\_bytes (number of bytes at this site), group, time\_create (when the replication request was made), percent\_files (percentage of files at the site), and percent\_bytes (percentage of bytes at the site).
\end{itemize}
\ \\
\emph{RequestList:} A list of requests.\\
Input options are:
\begin{itemize}
    \item \texttt{request}: request id~(*)
    \item \texttt{node}: name of the targeted site~(*)
    \item \texttt{dataset}: dataset name as a part of the request~(*)
    \item \texttt{block}: block name as a part of the request~(*)
    \item \texttt{requested\_by}: requester name~(*)
\end{itemize}
Output:
\begin{itemize}
    \item \texttt{id}: request id
    \item \texttt{time\_create}: time of the request creation
    \item \texttt{requested\_by}: requester name
    \item \texttt{list of sites}: for each site
     \item \texttt{node\_id}: target site id
    \item \texttt{name}: target site name
\end{itemize}

\end{appendices}

\bibliographystyle{spphys}
\bibliography{biblio}

\end{document}